\documentclass[12pt]{article}
\usepackage{graphicx}
\begin{document}
\title{On the fate of singularities and horizons\\ in higher derivative gravity}
\author{\normalsize Bob
Holdom\thanks{bob.holdom@utoronto.ca} \\
\small {\em Department of Physics, University of Toronto}\\\small {\em
Toronto, Ontario,}
M5S1A7, CANADA}\date{}\maketitle
\begin{picture}(0,0)(0,0)
\put(310,205){UTPT-02-09}
\put(310,190){hep-th/0206219}
\end{picture}
\begin{abstract}  
We study static spherically symmetric solutions of high derivative gravity theories, with
4, 6, 8 and even 10 derivatives.
Except for isolated points in the space of theories
with more than 4 derivatives, only solutions that are nonsingular near the origin are found.
But these solutions cannot smooth out the Schwarzschild singularity without the appearance of
a second horizon.
This conundrum, and the possibility of singularities at finite $r$,
 leads us to study numerical solutions of theories truncated at four derivatives.
Rather than two horizons we are led to the suggestion that
the original horizon is replaced by a rapid nonsingular transition from weak to
strong gravity. We also consider this possibility for the de Sitter horizon.
\end{abstract}
\baselineskip 18pt
\section{Introduction}
The vacuum Schwarzschild (Schd) solution of general relativity suffers from a 
singularity at $r=0$, but perhaps a more
meaningful observation is
that the Schd solution cannot be trusted in a finite region enclosing the origin.
At a nonzero radius inside the horizon a curvature invariant has grown as large as the Planck mass scale,
at which point the Einstein action should no longer provide the correct description.
 For example in the standard Schd
coordinate system ${R}_{\mu \nu \sigma \rho
}{R}^{\mu \nu \sigma \rho } = 48G^2M^2/{r}^{6}$,
and thus the Schd solution breaks down at a radius of order $r_s= (48G^4M^2)^{1/6}$.
For large $M$ this is much smaller than the horizon size $2GM$, but it is much larger than
the Planck length $\sqrt{G}$.

There have been discussions within general relativity on how the diverging curvatures
can be tamed \cite{g}. If a suitable matter distribution is postulated and allowed to
violate the strong energy condition,
then nonsingular black-hole-like solutions do arise. Although the
 curvatures in the core region are now finite they
are still characterized by the Planck scale, and thus the Einstein action can only
provide a crude description.  But an interesting aspect
 of such solutions is the appearance of
a second horizon, in the vicinity of $r_s$.

No matter what the theory is or what matter is present, it appears that nonsingular
spherically symmetric static solutions can never have only one horizon.
In the standard coordinate system the metric of interest is
\begin{equation}d{s}^{2}  =  -B(r)d{t}^{2}+A(r)dr^2+r^2(d\theta^2+\sin^2\theta\phi^2).
\label{met}\end{equation}
If the curvature invariants $R$, ${R}_{\mu \nu}{R}^{\mu \nu }$, ${R}_{\mu \nu \sigma \rho
}{R}^{\mu \nu \sigma \rho }$... are to be nonsingular at the origin, what does this imply
about the behavior of $A(r)$ and $B(r)$? For small $r$ we find that\footnote{In lieu of a proof we have
scanned through a variety of power law behaviors of $A(r)$ and $B(r)$ near the origin and a variety of
curvature invariants, similar to our scan for solutions in next section.}
\begin{eqnarray}A( r ) & =&  1+{a}_{2}{r}^{2}+...
\nonumber\\B( r ) & =& b(1+{b}_{2}{r}^{2}+...)
\end{eqnarray}
In this case $R=6(a_2-b_2)$ at $r=0$ and the higher invariants are polynomials in $a_2$ and $b_2$.
(The constant $b$ is affected by a rescaling of $t$ and can be ignored.)
The important point is that $A(0)$ is fixed to be unity. 
But inside the horizon of the Schd solution $A(r)$
is negative, a fact directly connected with the existence of the Schd horizon where the signs of
$A(r)$ and $B(r)$ change. Then to match onto a nonsingular solution with $A(0)=1$
another sign change is required. To preserve a time-like direction $B(r)$ must also
 change sign again. Thus $A(r)$ and $B(r)$ must each have an even number of
sign changes between $r=0$ and $r=\infty$, rather than the single sign 
change implied by a single horizon.

Given this result, a theory of gravity capable of describing Planck scale physics should address
the following three questions. Are these metrics that describe large but
nonsingular curvatures near the origin actually solutions to the theory? Do
solutions that are singular at the origin continue to be present, or are they banished
altogether? How does a nonsingular solution at the origin match onto a sensible solution at large $r$
while experiencing an even number of sign changes,
without encountering a singularity at a finite $r$?

To model Planck scale physics we consider theories where the Einstein
action is extended to include terms with more factors of the
curvature tensor and its covariant derivatives.
 We will explicitly study
spherically symmetric static solutions of these higher derivative theories truncated at
various orders in derivatives. In particular we will construct solutions as
a series expansion about the origin for theories with 6, 8 and even 10 derivatives.
And for the general theory truncated at four derivatives we will
be able to construct numerical solutions that interpolate
between the small curvature, weak gravity behavior at large $r$ and the
high curvature, strong gravity behavior at small $r$. Some aspects of our
results may be true at any order in the derivative expansion.

In the $2+4$ derivative theory (the theory with the Einstein term
and the general set of 4 derivative terms) the Schd solution is still present.
 Ref.~\cite{a} applied the series expansion approach to this
theory to find, in addition to the Schd solution, another class of singular solutions and a class of
nonsingular solutions. But it was not known which, if any, of these alternative solutions
near the origin could
match onto the desired weak gravity solution at large $r$. Our numerical work shall clarify
these issues.

Although the $2+4$ derivative theory has a variety of
 solutions near the origin, we find that only one class of solutions
remains for the typical action with more derivatives. This turns out to be
the nonsingular class of solutions, having the expected property $A(0)=1$. Such solutions are found
for any action considered. It is perhaps not surprising that nonsingular solutions exist, since
 spherically symmetric static solutions that are everywhere weak, produced
for example by a low density matter 
distribution, must be recovered. More surprising is that
\textit{singular} solutions, if they exist at all, only exist for very particular actions, 
of measure zero in the space of theories. (Among these very particular theories are those
that have received some attention. For example the Schd solution continues to exist in theories
where only powers of the curvature scalar $R$ appear in the action \cite{f}.)

This is not to say that solutions with singularities don't exist; the typical unphysical
solution would involve a curvature singularity at a finite $r$. But to find nonsingular
black-hole-like solutions our previous discussion indicates that we must search for solutions with several
sign flips. This appears difficult to
accomplish in higher derivative theories, either analytically or numerically, 
and it is not the focus of this work.
Rather we shall be more concerned with how a core region of strong gravity
matches onto a weak gravity large $r$ region
without any change of sign of metric components,
and thus with no horizons.

Before launching into the study of these solutions, it is natural to question the derivative expansion itself.
Of course a derivative expansion truncated at some finite order has problems with
physical interpretation; in particular it suffers from negative energy ghost modes when linearized.
But this problem is entangled with the effects of truncation and linearization; it 
is not fundamental in the sense that the underlying theory,
from which the derivative expansion is derived, should be quite sensible.
 Evidence of the role of nonlinearities in resolving the ghost problem
was found for a different class of metrics
of FRW form in \cite{c}, where ghost modes were found to decouple from positive energy matter
when the nonlinear solutions were considered.
In any case results that hold at arbitrary order in the nonlinear derivative expansion,
 such as the existence of nonsingular
solutions near the origin as described above, would appear to carry some significance.

The solutions near the origin and their implications are discussed in more
detail in the next section.
In sections 3 and 4 we turn to the general $2+4$ derivative theory and
 numerically analyze the non-Schd solutions that are respectively
singular and nonsingular near the origin.  We show how these solutions
match onto the desired weak gravity solutions at large $r$, and in particular that
they do so without encountering horizons.
 In section 5 we turn to the de Sitter space, which in static coordinates also has
a horizon at a finite radius. Here we again explore the idea that the horizon is replaced by a
boundary where a transition from weak to strong gravity occurs; now the picture is inside-out 
with weak (strong)
gravity on the inside (outside). We conclude in section 6.

\section{Solutions near the origin}
We consider a general gravitational action with any number of derivatives.
The field equations ${H}_{\mu \nu }^{ }  =  {T}_{\mu \nu }$ determining $A(r)$ and
$B(r)$ can be found by
substituting the metric (\ref{met}) into the action and varying with respect to $A(r)$ and $B(r)$;
this gives $H_{rr}$ and $H_{tt}$ respectively.
${H}_{\theta \theta }$ can then be found from the Bianchi identity \cite{a},
\begin{equation}{H}_{\theta \theta }  =  {\frac{{r}^{3}}{2}}( {\frac{{H}_{rr}}{A}} )'+{\frac{{r}^{2}{H}_{rr}}{A}}+{\frac{{r}^{3}B'{H}_{rr}}{4AB}}+{\frac{{r}^{3}B'{H}_{tt}}{4{B}^{2}}}.\end{equation}
We first look for nonsingular solutions in the presence of a smooth
matter distribution with finite energy and pressure.

For our purposes we define nonsingular metrics as those 
without curvature singularities,
and we have already said that such metrics must have
$A(0)=1$. It is interesting to see how
this emerges as a property of solutions to the field equations. When $A(0)$ and $B(0)$ are
nonzero and finite the leading terms in the field
equations for small $r$ behave like $1/r^n$ where $n$ is the maximum 
number of derivatives in the action. This
behavior arises from the terms with no derivatives of $A(r)$ and $B(r)$, which nevertheless arise
from $n$ derivative terms in the action.
The coefficient of the $1/r^n$ term in each equation is
proportional to a polynomial in $A(0)$, and these polynomials vanish iff $A(0)=1$.
This result is thus
intrinsically connected with the nonlinearity of the theory.

All the terms behaving like inverse powers of $r$ in the equations must vanish,
and this leads to solutions of the form
\begin{eqnarray}A( r ) & =&  1+{a}_{2}{r}^{2}+\sum\limits_{n  \ge  4}^{} {{a}_{n}r}^{n},
\nonumber\\B( r ) & =& 1+{b}_{2}{r}^{2}+\sum\limits_{n  \ge  4}^{} {{b}_{n}r}^{n}.
\label{e4}\end{eqnarray}
We find that solutions of this form exist for general actions,
and we have tested actions up to 10 derivatives.
 $a_2$ and $b_2$ are free parameters of these solutions and all the higher order coefficients
are determined by them and by the matter distribution.
If we assume the presence of a spherically symmetric matter distribution described by
the energy and pressure profiles $\rho(r)$ and $p(r)$, and if an equation of state is specified, then we
find that the series expansions of $A(r)$, $B(r)$, $\rho(r)$ and $p(r)$ are completely determined 
in terms of $a_2$, $b_2$ and $\rho(0)$. In this sense this is a 3 parameter family of solutions.
This holds for general actions
except for isolated points in theory space; one such exception is general relativity, 
where $a_2$ and $b_2$ are directly determined by $\rho(0)$ and $p(0)$.

We now consider the possibility of singular solutions.
To find other solutions near the origin we look for other cases where the leading terms in the $1/r$ expansion
of the field equations vanish for special choices of the $A(r)$ and $B(r)$ expansion parameters,
\begin{eqnarray}A( r )  &=&  {a}_{\alpha }{r}^{\alpha }+{a}_{\alpha +1}{r}^{\alpha +1}+...
\nonumber\\B( r ) &=&  b_\beta{r}^{\beta }+{b}_{\beta +1}{r}^{\beta +1}+...\end{eqnarray}
Note that both the leading and next-to-leading
parameters can appear in the leading terms in the field equations.
If such a case is found then the expansion can be tested as a solution at higher orders.
Since various powers of $r^\alpha$ and 
$r^\beta$ appear in the nonlinear field equations it would seem miraculous to find solutions with
 $\alpha$ and $\beta$ noninteger; in any case we have restricted the search to integer
$\alpha$ and $\beta$. Our search is also blind to solutions that are not amenable to a
series expansion at the origin.

We have performed a scan through a range of values of $\alpha$ and $\beta$
 for a range of actions. In the case of the general $2+4$ derivative theory
 the scan yields the three sets of solutions found
in \cite{a}: the nonsingular solutions in (\ref{e4}) with $(\alpha, \beta)=(0, 0)$, 
the Schd solution with $(\alpha, \beta)=(1, -1)$,
 and another set of singular solutions to be described below with $(\alpha, \beta)=(2, 2)$.
For actions with more derivatives only the nonsingular solutions remain, 
except for specialized actions where various terms are set arbitrarily to zero.

The result then is that the static spherically symmetric solutions of a typical high derivative theory
are not singular at the origin. If singularities occur, they occur at a nonzero radius.

 We stress again that the Schd solution is a generic solution only in
theories truncated at two derivatives (general relativity) or four derivatives. 
In typical theories with six or more
derivatives the Schd solution will remain only as an approximate weak gravity solution in the
large $r$ region where the Einstein term dominates. In fact by examining a series expansion
in $1/r$ of the field equations around the Schd solution
one finds (see also \cite{d}) that the corrections have the form,
\begin{eqnarray}A( r )  &=&  1/( 1-2GM/r )+{\cal O}( G^4M^2/{r}^{6} ),
\nonumber\\B( r ) & =&  1-2GM/r +{\cal O}( G^4M^2/{r}^{6} ).\label{e8}\end{eqnarray}
Once again we see that the Schd solution is a good approximation to an
exact solution down to a radius well
within the horizon, implying that $A(r)$ and $B(r)$ would have to change sign a second time
if they were to match onto (\ref{e4}). 
When the exact solution is extended in towards the origin
one must encounter a second horizon and/or a singularity at a finite radius.\footnote{The singularity
could be impassable. There is
also the possibility of a violent singularity at the origin of a type not amenable to a series expansion.
Note that some kind of singularity is expected in the case that $M$ is negative, since
a nonsingular negative energy solution would imply that Minkowski space is not stable \cite{e}.}

But there is a loophole in these arguments, related to the observation that an exact solution
of the form in (\ref{e8}) 
is not the unique solution in the weak gravity region \cite{a}.
As discussed in the next section, there are other solutions involving
 the massive modes in the theory, giving rise to solutions 
involving terms of the Yukawa potential type. Thus the real question is what happens
when a more general exterior solution, which is Newtonian plus Yukawa in form at large $r$, is continued
in towards $r=0$.

To shed light on these issues
 it would be best to make a direct analysis of a theory with six or more derivatives, 
where the only solutions are nonsingular at the origin.
But the complexity of those equations will constrain our present work to
a numerical study of the $2+4$ derivative theory.
We will investigate how the various solutions existing near the origin of the $2+4$ derivative 
theory join onto the weak-gravity large-$r$ solutions. We are once again
dealing with solutions of truncated theory in a region where the truncated theory is not justified,
but it will be of interest to compare and contrast these solutions to the Schd solution.

\section{Domains of strong gravity}
The action of the general $2+4$ derivative theory can be cast in the form
\begin{equation}
S  =   {1\over 16\pi G}\int_{}^{}{d}^{4}x\sqrt {-g}(R
+a{R}^{2}+b{R}_{\mu \nu }{R}^{\mu \nu }).\label{e1}\end{equation}
We first consider the \textit{linearized}
version of this theory, where the equations for $A(r)$ and $B(r)$ have
 five independent solutions \cite{a}. For a weak constraint on $a$ and $b$,
assumed here to be satisfied, four of the solutions
have exponential dependence. The two that grow exponentially with $r$ are
 discarded to satisfy boundary conditions at infinity. 
This leaves the Newtonian solution and two solutions that resemble Yukawa potentials.
For any physical matter distribution the Yukawa potential terms will
coexist with the Newtonian term in the exterior solution (an example is given in \cite{a}).
 One of the Yukawa potential terms has a repulsive sign,
reflecting the ghost-like nature of the massive mode. The Yukawa potentials are
 typically swamped by the Newtonian term and offer little chance of detection
in weak gravity solutions \cite{a,b}.

On the other hand the mere existence of the Yukawa potential terms 
for some matter distribution for which
linearized gravity is applicable proves that the exact Schd solution of the full $2+4$ derivative theory is not 
the true exterior solution \cite{b}. There is also no reason to expect
 it to be the relevant solution when there is sufficient matter density to
cause the Schd horizon to form. We must study other exact
solutions of the theory that take a Newtonian plus Yukawa form in the exterior region.

Such solutions can be markedly different from the Schd solution in the interior region.
In particular the region of high curvature and strong gravity can be larger
than in the Schd solution. The idea that a region of Planck scale curvature can be of macroscopic
size is not conceptually new since, as described in the introduction, this is already
is implied by the Schd solution. If strong gravity extends out to nearly the radius of the
would-be Schd horizon, then $A(r)$ and $B(r)$ could retain their weak 
gravity signs for all $r$, and a horizon need not exist.

This possibility turns out to be illustrated by the other class of
 solutions in the $2+4$ derivative theory that are singular near the origin.
 Close to the origin they have the form
\begin{eqnarray}A( r )  &= & {{a}_{2}r}^{2}+{{{a}_{3}{r}^{3}+{a}_{4}{r}^{4}+\sum\limits_{n  \ge  5}^{} {a}_{n}{r}^{n}}^{}}^{}
\nonumber\\B( r ) & = & b_2{r}^{2}+{{{\frac{{a}_{3}}{{a}_{2}}}b_2{r}^{3}+{\frac{6{a}_{2}{a}_{4}+{{a}_{3}^{2}+2{a}_{2}^{3}}^{}}{8{a}_{2}^{2}}}b_2{r}^{4}+\sum\limits_{n  \ge  5}^{} {b}_{n}{r}^{n}}^{}}^{}
\label{e2}\end{eqnarray}
We shall be concerned with $a_2,b_2>0$, where $b_2$ is sensitive to a rescaling of $t$.
The existence of these solutions up to $r^3$ was shown in \cite{a}.
We have found that these are 5 parameter solutions where the parameters
($a_2$, $a_3$, $a_4$, $a_5$, $b_5$) determine ($a_n$, $b_n$) for
$n\ge 6$, for a given action.
The leading behavior of the curvature invariants [$R$, ${R}_{\mu \nu }{R}^{\mu \nu}$, 
${R}_{\mu \nu \sigma \rho}{R}^{\mu \nu \sigma \rho }$] is
[$(22a_3a_2^3+10a_2a_3a_4-a_3^3-60b_5a_2^3+36a_5a_2^2)/(4a_2^4r)$, $12/(a_2^2{r}^{8})$,
$24/(a_2^2{r}^{8})$], to be compared with [0, 0, $\sim 1/r^6$] for the Schd solution.
We note here that
any nonsingular mass distribution only affects ($a_n$, $b_n$) for $n\ge10$.

It is an interesting coincidence that there are five parameters here, just as in
 the linearized gravity solutions. 
This lends support to the idea that full solutions exist which join together the strong gravity small $r$
solutions with the weak gravity large $r$ solutions.
A numerical verification is made somewhat difficult
 by the existence of exponentially growing solutions
at large $r$, and by the extremely singular behavior of the equations at small $r$ away
 from the exact solutions. In particular, a numerical analysis
based on the initial value problem
at $r=0$ is not feasible. Instead the following strategy was adopted.

At some finite $r_0$ in the 
\textit{weak}
gravity region initial conditions are chosen so as to deviate only very slightly from the Schd solution,
and the equations are numerically integrated for $r$ both larger and smaller than $r_0$.
The Yukawa potential modes are induced and they grow for decreasing $r$. This can cause
the solution to deviate significantly from the Schd solution around the would-be horizon region, and
rather than changing sign, both
$A(r)$ and $B(r)$ can stay finite and positive throughout this region and down to $r=0$.
The initial conditions at $r_0$ are then finely tuned so that $A(r)$ and $B(r)$ near $r=0$ take the form
of the known solution in (\ref{e2}) to ${\cal O}(r^4)$. 

Additional fine tuning is needed to remove
the unwanted modes that grow exponentially with $r$. But it is easier just to repeat the whole process
for a larger $r_0$ where
the required deviation from the Schd solution is smaller; then the amplitude of the unwanted modes is
smaller and a sensible numerical solution extends out to larger $r$. 
It appears that the extent to which one
can push this is a purely numerical limitation, and that these numerical results
are sufficient to demonstrate the existence of exact solutions of this type.

\begin{figure}
\begin{center}
\includegraphics[width=10cm]{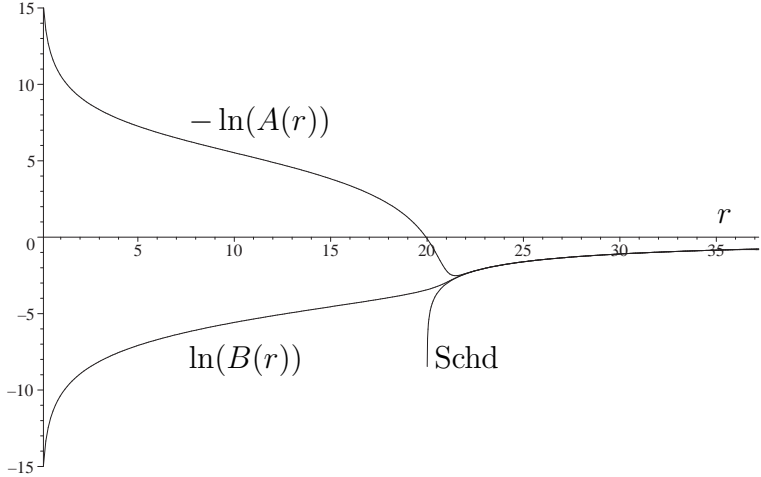}
\begin{picture}(0,0)(0,0)
\put(-20,95){$r$}
\put(-220,130){$-\ln(A(r))$}
\put(-220,40){$\ln(B(r))$}
\put(-127,40){Schd}
\end{picture}
\end{center}
\caption{A solution of the $2+4$ derivative theory with no horizon and a singularity at the origin, which
matches onto the Schd solution with ${\cal M}=10$ close to its horizon at $r=20$ (in units with
$G=1$). $A(r)$ and $B(r)$ exhibit an $r^2$ dependence near $r=0$.}\label{f1}
\end{figure}
We display one of these numerical vacuum solutions in Fig.~(\ref{f1}), where we plot the
functions $-\ln(A(r))$ and $\ln(B(r))$. These functions coincide
(after a suitable rescaling of $t$) with each other
and with the Schd solution
 in the weak gravity region, exterior to the would-be horizon.
 We have chosen $a=G/2$ 
and $ b=-G$ in (\ref{e1}) to simplify the
equations somewhat (corresponding, in the linearized theory, to
equal masses for the two massive modes), but the basic 
properties of the solutions are independent of this choice.
The mass ${\cal M}$ as deduced by the large $r$ behavior
 in this example is ${10}G^{-{\frac{1}{2}}}$. Larger ${\cal M}$ has also been 
considered and there
does not seem to be any limitation, other than numerical, to recover arbitrarily large mass solutions.

The new solutions have no horizon and a singularity at the origin. Their interior form,
 and the point where deviation from the Schd
solution occurs, is not uniquely determined by ${\cal M}$. This is as expected given the additional
parameters governing the strength of the Yukawa potentials in the exterior solution,
and the five parameters in (\ref{e2}).
We also find that $a_2\sim G^{-3}{\cal M}^{-4}$,
which is consistent with the curvature invariant
${R}_{\mu \nu \sigma \rho}{R}^{\mu \nu \sigma \rho }$ being of Planck size 
close to the would-be horizon, even for a large mass object.
Of course this is how the usual arguments for the existence of a horizon are avoided, since the
uniqueness of the Schd solution relies on small curvatures and the applicability of general relativity
down to radii well within the horizon.

This solution is intriguing, but it is occurring in a theory arbitrarily truncated in derivatives, and
the solution involves a region of high curvature where higher derivative terms
would be important. On the
other hand the Schd solution suffers from exactly the same problems. As in that case one may want to
presume that the higher derivative terms being ignored would serve to smooth out the singularity
at the origin while retaining the main qualitative feature, the fact that the transition between
weak and strong gravity has moved out to the would-be horizon radius.

\section{Nonsingular solutions}
We shall now turn to the nonsingular solutions of the $2+4$ derivative theory, where
additional insights will emerge.\footnote{We keep in mind though that these solutions need not
be truly representative of the nonsingular solutions in theories with even more derivatives.}
 A spherically symmetric nonsingular matter distribution will act
as a source for these solutions. We choose to study incompressible matter, where
 $\rho(r)$ will be given a fixed profile and the theory will determine the
required pressure $p(r)$. In particular we take $\rho(r)=\rho_0 \exp(-r^2/{\cal R}^2)$. 

\begin{figure}
\begin{center}
\includegraphics[width=10cm]{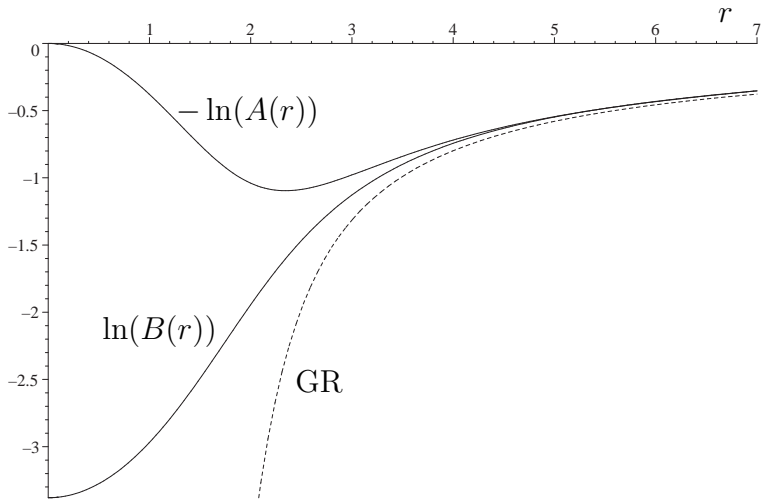}
\begin{picture}(0,0)(0,0) 
\put(-20,180){$r$}
\put(-225,143){$-\ln(A(r))$}
\put(-253,60){$\ln(B(r))$}
\put(-180,40){GR}
\end{picture}
\end{center}
\caption{A nonsingular solution of the $2+4$ derivative theory for the smooth mass distribution
with ${\cal R}=4/3$ and $M=1.1$ in Planck units.
 For the same distribution the dotted line shows $\ln(B(r))$ from general
relativity, which yields vanishing $B(r)$ and infinite pressure at some radius.}\label{f2}
\end{figure}
The same strategy that was used in the previous section
is less successful for the nonsingular solutions.
 Presumably this is related to the fewer number of parameters present in the nonsingular solutions
near the origin; compare (\ref{e4}) to (\ref{e2}).
Thus far we are only successful by taking ${\cal R}$ of order the Planck length $\sqrt{G}$,
but this seems to be purely a limitation of the numerical strategy which we now describe.
Noting that equations cannot be directly integrated from $r=0$,
we use the series expansion in (\ref{e4}) to ${\cal O}(r^{12})$ and express all
coefficients in terms of $a_2$, $b_2$, $p_0$ and $\rho_0$. This is
used to provide the initial conditions for the differential equations at a small finite value of $r$,
where the series solution can be trusted.
The equations can then be numerically integrated out from this point to large $r$. For a given
$\rho_0$, a tedious search of the ($a_2$, $b_2$, $p_0$) parameter space is then needed to match
 onto the weak gravity exterior solution, with the additional constraint that 
$p(r)\rightarrow 0$ for $r\gg {\cal R}$. We display
 one of these solutions in Fig.~(\ref{f2}); our choice here and in the following is ${\cal R}=4\sqrt{G}/3$.

\begin{figure}
\begin{center}
\includegraphics[width=10cm]{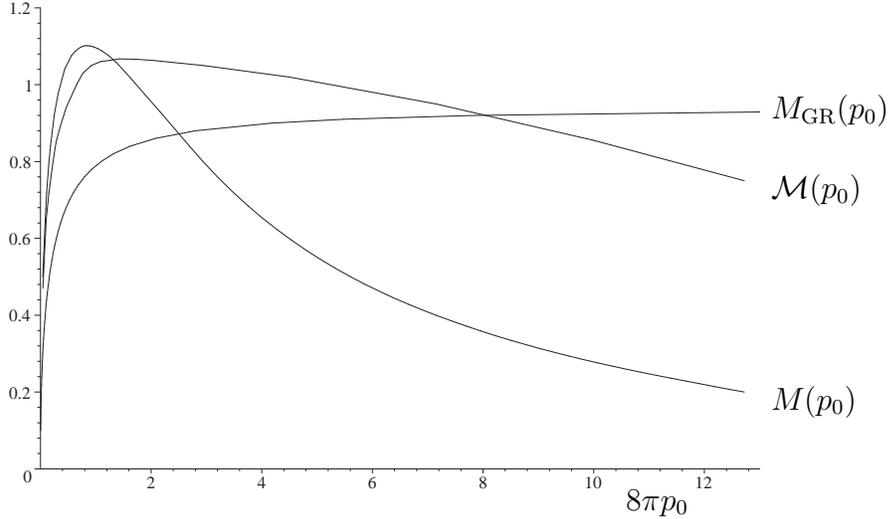}
\begin{picture}(0,0)(0,0) 
\put(0,140){$M_{\rm GR}(p_0)$}
\put(0,30){$M(p_0)$}
\put(0,110){${\cal M}(p_0)$}
\put(-55,-8){$8\pi p_0$}
\end{picture}
\end{center}
\caption{The mass in a fixed volume versus the pressure $p_0$ at the origin.
$M$ is the volume integral of the matter energy density and ${\cal M}$ is the mass
deduced from the exterior solution. Both definitions apply to the mass
in general relativity, $M_{\rm GR}$.}\label{f3}
\end{figure}
With the usual definition of mass $M  \equiv  \int_{}^{}\rho ( r )4\pi {r}^{2}d{r}$ we
have $M\propto\rho_0$ for fixed ${\cal R}$.
In general relativity the pressure at the origin $p_0$ increases with increasing $M$,
and $p_0$ becomes
infinite at a finite $M$. The whole notion of gravitational collapse is based on this basic
behavior, but we find a qualitatively different behavior in the $2+4$ derivative theory.
We refer to our derived $M(p_0)$ in Fig.~(\ref{f3}) where for
 small $p_0$ and $M$, $p_0$ increases with $M$, but more slowly than in 
GR. The higher derivative terms are acting to reduce the tendency to collapse,
and $p_0$ is still finite at the
particular mass that first causes infinite $p_0$ in GR.
$p_0$ continues to increase for increasing $M$ until the strong gravity effects cause a
peculiar phenomenon; a maximum value $M_{\rm max}$ occurs at a finite $p_0$.  
After this point $p_0$ continues to increase while $M$ \textit{decreases}!

Connected with this behavior is the fact that, unlike GR, the physical mass $\cal{M}$ 
of the object as deduced by the exterior solution is not
the same as $M$, the volume integral of $\rho(r)$.
From Fig.~(\ref{f3}) we see that
 $\cal M$ is smaller than $M$ for small $p_0$, while the reverse happens for sufficiently large
$p_0$. In the latter case strong gravity contributions are responsible for an increasing
fraction of the total $\cal M$, which decreases much less rapidly than $M$ for increasing $p_0$. 
We note that the maximum value $M_{\rm max}$ is achieved when
 the scalar curvature $R$ at the origin switches sign, being positive (negative) on the low (high)
pressure side of $M_{\rm max}$. The example in Fig.~(\ref{f2}) corresponds to
 $M\approx M_{\rm max}$.

Even though the $M(p_0)$ curve  in Fig.~(\ref{f3}) is
derived for an incompressible material, it has implications for massive objects with more physical equations 
of state. For the more standard
 part of the curve where $p_0$ is growing with $\rho_0$ (remember $\rho_0\propto M$), 
$p_0/\rho_0$ is also
growing. Thus when more matter is added to a given volume the object may or may not be able to
support the required increase in pressure, depending on its equation of state. If not,
gravitational collapse ensues. On the other part of the curve, $p_0$ \textit{decreases} with increasing
 $\rho_0$. Thus adding more matter to a given volume will tend to decrease the required pressure,
and the required $p_0/\rho_0$ decreases even more. Then the matter exerts too much
 pressure and the object will tend
to expand in size rather than contract. Thus the usual tendency for gravitational collapse is
absent on this part of the curve. The crossover between the two types of behavior
occurs at the value of $p_0/\rho_0$ at $M_{\rm max}$; for our example with ${\cal R}=4\sqrt{G}/3$
this is 0.39.

It remains to be checked whether this basic picture survives for much larger ${\cal R}$.
The question is whether a $M(p_0)$ curve corresponding to a larger volume, 
which would lie above the one 
in Fig.~(\ref{f3}), has the same general shape. 
But our present numerical approach is not powerful enough to deal with this question.

Finally, the concept of a maximum mass at a finite pressure leads to the following question.
 Suppose we start with the maximum mass configuration; what happens if the
mass in the fixed volume is increased still further? As in general relativity, we may expect
that a curvature singularity should develop at the origin. If so then the solution should transform
into one of the solutions of the previous section;
we have already mentioned
that the presence of a nonsingular mass distribution would only affect those solutions
at order ${\cal O}(r^{10})$. By continuity it seems appropriate that the new configuration would maintain
a finite $p_0$. In this way it appears that the ``maximum mass at finite pressure'' property of
 the nonsingular solution is related to the existence of the singular class of 
solutions. Unlike general relativity, the transition from a nonsingular solution to a singular
one does not lead to the appearance of a horizon.

Then what happens when there are 6 or more derivatives in the action, and there are
no singular solutions at the origin. Does this imply that the maximum mass of the
 $M(p_0)$ curve extends out to infinite $M$? This would be the most intriguing possibility;
the only nonsingular alternative to the horizon-free solutions discussed here
is the possibility of solutions with at least two horizons.

\section{The de Sitter horizon}
Another horizon of physical interest occurs in the de Sitter metric, which in
the static coordinate system has
\begin{equation}B(r)=1-{\Lambda\over 3}r^2\;\;\;,\;\;\;A(r)=1/B(r).
\label{e5}\end{equation}
The physical region is interior to the horizon at $r=\sqrt{3/\Lambda}$ and for small
 cosmological constant it is a weak gravity,
small curvature region with $R=4\Lambda$. 
But we may wonder whether
the weak gravity region can border on a strong gravity region in such a way that there
is no horizon, with the border being roughly
 the location of the would-be horizon. We will find that actions containing terms
with six or more derivatives are necessary 
to deal with this question. Thus we must rely less on numerical analysis and more on series expansions,
 both in the weak and strong gravity regions. We find results that
are consistent with the idea of horizon elimination.

First we note that the de Sitter metric when expanded about the origin is a
member of the nonsingular class of solutions in (\ref{e4}). 
We are interested in other solutions of this class where the scalar curvature $R$
is close to being constant, departing significantly from $4\Lambda$
 only when the would-be horizon is approached.
We can write such solutions to theories with a cosmological constant and no matter in the form
\begin{eqnarray}A(r)&=&1/(1-({\Lambda\over 3}-\varepsilon)r^2)+\sum\limits_{i  \ge  2}{a}_{2i}r^{2i},
\nonumber\\B(r)&=&1-({\Lambda\over 3}+\varepsilon)r^2+\sum\limits_{i  \ge  2} {b}_{2i}r^{2i}.\label{e10}\end{eqnarray}
The parameter $\varepsilon$ characterizes the deviation from the de Sitter metric.
The coefficients $a_i$ and $b_i$ vanish with
$\varepsilon$ and they also depend on $\Lambda$ and the action being considered. 

For the $2+4$ derivative action we find that solutions with a nonzero
$\varepsilon$ do not in fact produce a departure of $R$ from $4\Lambda$. Actions with more
derivatives are required for this to happen; in addition such actions cause the 
$\Lambda$ in (\ref{e10})
to shift by an amount of order $\Lambda^3$ from the cosmological constant in the action.
When this is taken into account we find that an $\varepsilon$ expansion gives\footnotemark[3]
\begin{equation}R=4\Lambda+\varepsilon^2f_\Lambda(r^2)+...\label{e11}\end{equation} 
$f_\Lambda(r^2)$ has a nonvanishing $\Lambda=0$ limit and $f_\Lambda(0)=0$.
 By taking $\varepsilon$ appropriately small we can
suppose that $|R-4\Lambda|$ will remain small until $r$ approaches the would-be horizon.
Numerically we find that $\varepsilon$ has to be exponentially small;
this is reminiscent of the exponentially small departure from the Schd solution in the exterior region
discussed in the previous sections.

To explore the large $r$ strong gravity region we expand the equations in powers of $1/r$.
We find a new one-parameter family
of solutions that can only arise when six or more derivative terms appear in the action.
These new solutions are typically a corrected form of the anti-de-Sitter-Schd metric,
since pure AdS-Schd solutions only occur for specialized actions.
The appearance of corrections are not unexpected given the corrected Schd metric
in (\ref{e8}). The leading corrections are deduced by 
solving the equations to the appropriate order in $1/r$, and we find\footnote{
Actions with 6 and 8 derivatives were considered.}
\begin{eqnarray}A(r)&=&1/(1+{\hat{\Lambda}\over 3}r^2-2GM/r)+{\hat a}_8G^5M^2/r^8+...\nonumber\\
B(r)&=&1+{\hat{\Lambda}\over 3}r^2-2GM/r+{\hat b}_6G^4M^2/r^6+...\end{eqnarray}
These are strong gravity solutions since the positive $\hat{\Lambda}$ is fixed by the theory
to be Planckian in size, $\hat{\Lambda}\approx1/G$.
The dimensionless ${\hat b}_6$ and ${\hat a}_8$ constants are also determined by the action
being considered,
whereas $M$ is the free parameter.

If the corrections were absent then the scalar curvature would be $-4\hat{\Lambda}$
and there would be a horizon at $r=(6GM/\hat\Lambda)^{1\over3}
\approx G^{2\over3}M^{1\over3}$ for large $M$.
But at large $r$ the corrections modify the scalar curvature as follows,
\begin{equation}R=-4\hat{\Lambda}-{2{\hat a}_8\over3}{G^5\hat{\Lambda}^2M^2\over r^6}+...
\label{e12}\end{equation}
Thus in the vicinity of the would-be horizon the departures from the pure AdS-Schd metric
are becoming large, and so what actually happens there is not known. Note that
 this modification of the horizon region has emerged automatically from the general solution.
 The one parameter
$M$ can then be adjusted so that the radius of this would-be horizon 
is in the vicinity of the would-be horizon of the weak gravity
interior solution.
In both the interior and exterior regions $A(r)$ and $B(r)$ are positive,
and so the matching of solutions can occur without a horizon.

These results are of course not sufficient to prove the
existence of an exact solution that joins together the small and large $r$ behaviors we have described,
but the results are consistent with such a solution.
A numerical analysis of this horizon-free possibility remains to be performed.
The resulting picture is of a large mass object in a 
strong AdS-space that has as its interior a weak dS-space. The ``mass of the universe''
$M\approx G^{-2}\Lambda^{-{3\over2}}$ is enormous, since it is of order a Planckian mass
density times the volume of the interior space.

\section{Conclusion}
We have studied static spherically symmetric
solutions of higher derivative gravity, and have found that when there are six or more derivatives
the solutions are typically nonsingular near the origin. But the nature of these solutions implies that the
complete solutions with Newtonian large $r$ behavior must have an even number
of horizons. Our study of $2+4$ derivative gravity provides some insights into how 
a solution with a strong gravity core region can realize the zero horizon option.
The region of strong gravity extends out to the radius of the would-be
horizon, which then negates the usual arguments for the existence of a horizon. We
exhibited horizon-free solutions of this type that were singular
at the origin. It remains to show how this picture
persists in the higher derivative theories where solutions are always nonsingular at the origin.

We also discussed some unexpected properties of nonsingular solutions in $2+4$ 
derivative gravity that affect the
inevitability of gravitation collapse. A high pressure accumulation of
matter can have the property that increasing energy density corresponds to decreasing pressure. Associated
with this is a maximum for the amount of matter in the given volume.
This sounds familiar, but in this case horizons and black holes play no role. Rather, the maximum mass
configuration corresponds to a particular finite pressure at the origin,
 and this configuration can be approached
from both the low pressure and high pressure sides. Numerical limitations confined our
study of this phenomena to Planck mass objects, and so it remains to extend this picture to
much larger masses.

It could be expected that strong gravity and higher derivatives would help to resolve
singularities. But we have seen that a transition region from weak to strong gravity can
also take the place of a horizon. Of most interest are
solutions that are nonsingular and same-sign everywhere; such solutions have
to be investigated in theories with six or more derivatives to see
whether they exist for smooth mass distributions of arbitrary energy density.  
We have seen in the previous section that theories with six or more derivatives
are necessary to describe the possible elimination of the de Sitter horizon. In both the Schd and de Sitter
cases the deviations in the weak gravity regions are exponentially small until the would-be
horizon is reached.
It remains to be seen whether regions of strong gravity can be intruding on our weak
gravity universe.

\section*{Acknowledgments}
I thank E. Poppitz for discussions. This research was supported in part by the Natural Sciences 
and Engineering Research Council of Canada.

\end{document}